\documentclass[a4paper,11pt]{article}

\usepackage{latexsym}
\usepackage{amsfonts} 
\usepackage[mathscr]{euscript}
\usepackage{amsmath,mathrsfs,bm,amssymb,color}

\title{\textsf{
Nagaoka's theorem in the Holstein-Hubbard model
}}

\date{\empty}
\author{
Tadahiro Miyao\\ 
 {\it Department of Mathematics,}
{\it Hokkaido University,}\\
{\it Sapporo 060-0810, Japan}\\
E-mail:
 miyao@math.sci.hokudai.ac.jp
}

\newcommand{\one}{{\mathchoice {\rm 1\mskip-4mu l} {\rm 1\mskip-4mu l}
{\rm 1\mskip-4.5mu l} {\rm 1\mskip-5mu l}}}
\newcommand{\h}{\mathfrak{H}}

\newcommand{\Fock}{\mathfrak{F}}

\newcommand{\ran}{\mathrm{ran}}

\newcommand{\la}{\langle}
\newcommand{\ra}{\rangle}

\newcommand{\BbbR}{\mathbb{R}}
\newcommand{\BbbN}{\mathbb{N}}
\newcommand{\BbbZ}{\mathbb{Z}}

\newcommand{\BbbC}{\mathbb{C}}
\newcommand{\vepsilon}{\varepsilon}
\newcommand{\vphi}{\varphi}

\newcommand{\Nf}{N_{\mathrm{f}}}
\newcommand{\Hf}{H_{\mathrm{f}}}

\newcommand{\Cone}{\mathfrak{P}_M}
\newcommand{\cone}{\mathfrak{P}}

\newcommand{\no}{\nonumber \\}

\newcommand{\bphi}{{\boldsymbol \phi}}

\newcommand{\HH}{H_{\mathrm{H}}}

\newcommand{\Ue}{U_{\mathrm{eff}}}

\newcommand{\bsig}{{\boldsymbol \sigma}}
\newcommand{\btau}{\boldsymbol \tau}
\newcommand{\Up}{{\boldsymbol \Uparrow}}

\newcommand{\bfx}{x}
\newcommand{\bfy}{y}

\newcommand{\bfA}{A}
\newcommand{\bfr}{r}
\newcommand{\bq}{{\boldsymbol q}}

\begin{document}

\newtheorem{define}{Definition}[section]
\newtheorem{Thm}[define]{Theorem}
\newtheorem{Prop}[define]{Proposition}
\newtheorem{lemm}[define]{Lemma}
\newtheorem{rem}[define]{Remark}
\newtheorem{assum}{Condition}
\newtheorem{example}{Example}
\newtheorem{coro}[define]{Corollary}

\maketitle
\begin{abstract}
Nagaoka's theorem on ferromagnetism in the Hubbard model is extended to
 the Holstein-Hubbard model. This shows that Nagaoka's ferromagnetism is
 stable  even if the electron-phonon interaction is taken into account.
We also prove that Nagaoka's ferromagnetism is stable under the
 influence of the quantized radiation field.
\end{abstract}

\section{Introduction}

To build  rigorous theory of ferromagnetism is  a challenging problem.
The Hubbard model is one of the most fundamental model for ferromagnetic
metals.
Nagaoka constructed  a first  rigorous example of the ferromagnetism in
this model \cite{Nagaoka}.
He proved that the ground state of the model exhibits ferromagnetism
when  one electron is fewer than half-filling 
and the Coulomb strength $U$ is infinitely large. 
We remark that Thouless also disscused the same mechanism in \cite{Thouless}.
Since their discoveries, there have been several crucial developments
\cite{Lieb,Mielke, Tasaki3},
however,  Nagaoka's theorem   has been a major milestone in this field.
There are several studies concerning  Nagaoka's theorem;
a generalized version of the theorem was given by Tasaki \cite{Tasaki1};
Shastry et al. studied  the instability of Nagaoka's ferromagnetic state
\cite{SKA}, while 
 Kohno extended  the theorem to the Hubbard ladders  with several holes \cite{Kohno}.
This theorem still provides  attractive field of studies, see,  e.g.,
\cite{KT, KSV}.

On the one hand, electrons always interact with phonons in actual metals,
on the other hand, ferromagnetism is experimentally  observed in
various metals and has a
wide range of uses in daily life.
Therefore, if  Nagaoka's theorem contains an essence of  real ferromagnetism,
 Nagaoka's ferromagnetism should be  stable under the influence of the electron-phonon interaction. 
The main purpose of this paper is to prove this stability.
In addition, we show that   stability holds  even if 
 the electrons interact with the quantized radiation field.
To prove these results, we apply  operator theoretic correlation inequalites studied by Miyao 
in the previous works \cite{Miyao1, Miyao8, Miyao2, Miyao4, Miyao5,
Miyao6, Miyao7}.

We mention some related works. In \cite{Lieb}, Lieb gave an example of
ferromagnetism in the Hubbard model at half-filling. Recently, Lieb's 
ferromagnetism  is shown to be stable even if the electron-phonon
interaction is not so strong \cite{Miyao9}.
This result is consistent with our results in the present paper .

The organization of the present paper is as follows.
In Section \ref{Results}, we first review Nagaoka's theorem on the Hubbard model.
Then we state stability theorems on the Holstein-Hubbard model and 
Hubbard model coupled to the quantized radiation field. 
In Sections \ref{Pf1}-\ref{PFRad}, we prove main theorems.
In Appendix \ref{Tools}, we prove some correlation inequalties which play an important role in the proofs.

\section{Results}\label{Results}

\subsection{Nagaoka's theorem revisited}
First of all, we review Nagaoka\rq{}s theorem  in the
Hubbard model.
The Hamiltonian of the Hubbard model is 
\begin{align}
\HH=&\sum_{x,y\in \Lambda}
\sum_{
\sigma\in \{\uparrow, \downarrow\}}t_{xy}
 c_{x\sigma}^*c_{y\sigma}
+U\sum_{x\in \Lambda} n_{x\uparrow} n_{x\downarrow}+
\sum_{{x, y\in
 \Lambda}\atop{ x\neq y}}U_{xy}n_{x}n_y.
\end{align} 
$\HH$ acts in the $N$-electron Hilbert space
$\mathfrak{E}_{\mathrm{el}}=\wedge^N(\ell^2(\Lambda) \oplus \ell^2(\Lambda))$,
where $\wedge^n \mathfrak{h}$ is the $n$-fold anti-symmetric tensor
product of $\ell^2(\Lambda) \oplus \ell^2(\Lambda)$.
$c_{x\sigma}^*$ and $c_{x\sigma}$ are the standard fermionic operators
which create and annihilate the electron at site $x$ with spin $\sigma$.
$n_{x\sigma}=c_{x\sigma}^*c_{x\sigma}$ is the electron number operator
at site $x$ with spin $\sigma$. Moreover, $n_x=n_{x\uparrow}+n_{x\downarrow}$.
$t_{xy}$ is the hopping matrix element. $U$ and $U_{xy}$ are the local
and nonlocal Coulomb matrix elements, respectively.

In what follows, we assume the following:

\begin{flushleft}
{\bf (A. 1)} $\{t_{xy}\}$ and $\{U_{xy}\}$ are real symmetric matricies.
\medskip\\

{\bf (A. 2)} $t_{xy} \ge 0$ for all $x, y\in \Lambda$.
\medskip\\

{\bf (A. 3)} $N=|\Lambda|-1$.
\end{flushleft}

We derive an  effective Hamiltonian describing the system with $U=\infty$.
Let $P$ be the  Gutzwiller projection given by 
\begin{align}
P=\prod_{x\in \Lambda} (\one-n_{x\uparrow}n_{x\downarrow}),
\end{align} 
where $\one $ denotes the identity operator.
$P$ is the orthogonal projection onto
the subspace with
no doubly occupied sites.

\begin{Thm}
Assume {\bf (A. 1)}, {\bf (A. 2)} and {\bf (A. 3)}.
We define the effective Hamiltonian $H_{\mathrm{H}, \infty}$ by $H_{\mathrm{H}, \infty}=P\HH^{U=0}
 P$, where  $\HH^{U=0}$ is the Hamiltonian $\HH$ with $U=0$.
For all $z\in \BbbC\backslash \BbbR$, we have 
\begin{align}
\lim_{U\to \infty}(\HH-z)^{-1}=(H_{\mathrm{H}, \infty}-z)^{-1} P
\end{align} 
in the operator norm topology.
\end{Thm} 

Set $\mathfrak{E}:=\ran(P)$,
 where $\ran(P)$ is the range of $P$.
We  denote the restriction of   $H_{\mathrm{H},
\infty}$ to   $\mathfrak{E}$ by the same symbol.

To state Nagaoka\rq{}s thereom, we need some preparations.
The set of spin configurations  with a single hole  is denoted by $\mathcal{S}$:
 \begin{align}
\mathcal{S}=
&\Big\{
\bsig=\{\sigma_x\}_{x\in \Lambda}\in \{\uparrow, 0,  \downarrow\}^N\, \Big|\, \no
& \hspace{2cm}\mbox{There exists an $x_0$ such that $\sigma_{x_0}=0$ and 
$\sigma_x\neq 0$ if
 $x\neq x_0$}
\Big\}.\label{Conf}
\end{align} 
We say that 
the $x_0$ in (\ref{Conf}) is  the {\it  position  of the hole}, and
$\{\sigma_x\}_{x\neq x_0}$ is  the {\it  spin configuration of electrons. } 
For each $\bsig\in \mathcal{S}$, the position of the hole is denoted
by $x_0(\bsig)$.

For each $\bsig\in \mathcal{S}$, we denote the number of up spins
(resp.,  down spins) in $\bsig$ by
$n_{\uparrow}(\bsig)$ (resp.,  $n_{\downarrow}(\bsig)$), respectively.
For each $M\in \{-(|\Lambda|-1)/2, -(|\Lambda|-3)/2, \dots, (|\Lambda|-1)/2\}$, 
we set $\mathcal{S}_M=\{\bsig\in \mathcal{S}\, |\, n_{\uparrow}(\bsig)-n_{\downarrow}(\bsig)=2M\}$.

An element   $(x, \bsig)$ in $\Lambda \times \mathcal{S}_M$ is called 
the 
{\it
 hole-spin configuration},  if  $x=x_0(\bsig)$. In
 what follows,  we denote the set of all hole-spin configurations by
 $\mathcal{C}_M$.

Let $(x, \bsig)\in  \mathcal{C}_M$.
For each  $ y\in \Lambda\backslash \{x\}$, we define a map $S_{yx}:\, \mathcal{C}_M
\to \mathcal{C}_M \cup \{\emptyset\}$ by 
\begin{align}
S_{yx}(x, \bsig)=(y, \bsig'), \label{DefSxy}
\end{align} 
where 
$
\bsig'=
\{\sigma_z'\}_{z\in \Lambda}
\in \mathcal{S}_M
$ is defined by
\begin{align}
\sigma_z'=\begin{cases}
\sigma_y & \mbox{if $z=x$}\\
0 & \mbox{if $z=y$}\\
\sigma_z & \mbox{otherwise}
\end{cases}
\end{align} 
 and $S_{yx}(z, \bsig'')=\emptyset$ for all $(z, \bsig'') \in
 \mathcal{C}_M$ with  $z\neq x$.

\begin{define}[Connectivity condition]\label{HC}
{\rm 
We say that $\Lambda$ has 
the {\it  connectivity}, 
if the following condition holds:
For every pair of hole-spin configurations $(x, \bsig), (y, \btau)\in
 \mathcal{C}_M$, there exist sites $
x_1, \dots, x_{\ell}\in \Lambda
$ with $x_1=x,\ x_{\ell}=y$  such that 
\begin{align}
t_{x_{\ell}x_{\ell-1}} t_{x_{\ell-1} x_{\ell-2}}\cdots t_{x_2x_1}\neq 0
\end{align} 
and 
\begin{align}
\Big(
S_{x_{\ell} x_{\ell-1}}  \circ 
S_{x_{\ell-1} x_{\ell-2}}
\circ
\cdots \circ
S_{x_2x_1}
\Big)(x, \bsig)=(y, \btau). \ \ \ \ \ \  \diamondsuit
\end{align} 
}
\end{define} 

The total spin $S$ is an important quantity in the present paper.
The spin operators are given by 
\begin{align}
S^{(3)}=\frac{1}{2}\sum_{x\in \Lambda}(n_{x\uparrow}-n_{x\downarrow}),\ \
 \ 
 S^{(-)}=\sum_{x\in \Lambda} c_{x\downarrow}^* c_{x\uparrow},\ \ \
 S^{(+)}=
\big(S^{(-)}\big)^*. \label{DSpin}
\end{align} 
They all commute with the Hamiltonian $H_{\mathrm{H}}$.
The total spin operator is defined by 
\begin{align}
\big(
S_{\mathrm{tot}}
\big)^2=
\big(
S^{(3)}
\big)^2+\frac{1}{2}
S^{(+)}S^{(-)}+\frac{1}{2} S^{(-)} S^{(+)} \label{DefSpin}
\end{align} 
with eigenvalues $S(S+1)$.
We are interested in the eigenvalues of $\big(
S_{\mathrm{tot}}
\big)^2$ for the ground states.

The following theorem is due to Tasaki \cite{Tasaki1}.

\begin{Thm}[Generalized Nagaoka's theorem]\label{Hubbard}
Assume {\bf (A. 1)}, {\bf (A. 2)} and {\bf (A. 3)}.
 Assume that $\Lambda$ satisfies the  connectivity condition.
Then the  ground state of $H_{\mathrm{H}, \infty}$ has total spin  $S=(|\Lambda|-1)/2$ and is
 unique apart from  the trivial $(2S+1)$-degeneracy. 
\end{Thm}

\begin{rem}{\rm
In \cite{Tasaki2}, the sufficient condition for the connectivity  is given.
Using the condition, we know that models with the following (i) and (ii)
satisfy the connectivity condition:
\begin{itemize}
\item[(i)]
  $\Lambda$ is  a triangular, square cubic, fcc, or  bcc lattice;
  \item[(ii)]   $t_{xy}$ is   nonvanishing  between nearest neighbor sites.
\end{itemize}
}
\end{rem}

\subsection{Stability I: The Holstein-Hubbard model}
The Hamiltonian of the Holstein-Hubbard model is given by 
\begin{align}
H=&\sum_{x,y\in \Lambda}
\sum_{
\sigma\in \{\uparrow, \downarrow\}}t_{xy}
 c_{x\sigma}^*c_{y\sigma}
+U\sum_{x\in \Lambda} n_{x\uparrow} n_{x\downarrow}+
\sum_{{x, y\in
 \Lambda}\atop{ x\neq y}}U_{xy}n_{x}n_y\no
&+\sum_{x, y\in \Lambda}
g_{xy}n_{x}(b_y^*+b_y)+\sum_{x\in \Lambda} \omega b_x^*b_x. \label{ExtendedHH}
\end{align} 
$H$ acts in the Hilbert space $\mathfrak{E}_{\mathrm{el}} \otimes \Fock$. Here, 
$\Fock$ is the bosonic Fock space over $\ell^2(\Lambda)$ defined by 
$\Fock=\oplus_{n=0}^{\infty} \otimes_{\mathrm{s}}^n \ell^2(\Lambda)$,
where $\otimes_{\mathrm{s}}^n \mathfrak{h}$ is the $n$-fold symmetric
tensor product of $\mathfrak{h}$.
$b_x^*$ and $b_x$ are the bosonic creation- and annihilation operators
at site $x$.  
$g_{xy}$ is the strength of the electron-phonon interaction.
The phonons are assumed to be dispersionless with energy $\omega>0$.
$H$ is a self-adjoint operator, bounded from below.

We assume the following:
\begin{flushleft}
{\bf(A. 4)} $\{g_{xy}\}$ is a real symmetric matrix.
\end{flushleft}

\begin{Thm}\label{HHEff}
Assume {\bf (A. 1)}, {\bf (A. 2)}, {\bf (A. 3)} and {\bf (A. 4)}.
We define the effective Hamiltonian  $H_{\infty}$ by  $
H_{\infty}=
PH^{U=0}
P$, where  $H^{U=0}$ is  the Hamiltonian $H$ with $U=0$.
For all $z\in \BbbC\backslash \BbbR$, we have
\begin{align}
\lim_{U\to \infty}(H-z)^{-1}=(H_{\infty}-z)^{-1} P
\end{align} 
in the operator norm topology.
\end{Thm}  

We denote the restriction of $H_{\infty}$ to 
$\mathfrak{E}\otimes \Fock$ by the same symbol.

\begin{Thm}\label{Holstein}
Assume {\bf (A. 1)}, {\bf (A. 2)}, {\bf (A. 3)} and {\bf (A. 4)}.
 Assume that $\Lambda$ satisfies  the   connectivity condition.
Then the  ground state of $H_{\infty}$ has total spin $S=(|\Lambda|-1)/2$ and is
 unique apart from  the trivial $(2S+1)$-degeneracy. 
\end{Thm}

\subsection{Stability II: The Hubbard model coupled to the quantized
  radiation field}
We consider an  $N$-electron system coupled to the quantized radiation
field. 
Suppose that 
the lattice $\Lambda$  is embedded into the region $V=[-L/2,
L/2]^3\subset \BbbR^3$
with $L>0$.
The system is described by the following Hamiltonian:
\begin{align}
\mathsf{H}=&\sum_{ {x, y\in \Lambda}\atop{ \sigma=\uparrow, \downarrow}}
t_{xy}
\exp\Bigg\{
i   \int_{C_{\bfx\bfy}} d\bfr\cdot  \bfA(\bfr)\Bigg\}c_{\bfx\sigma}^*
 c_{\bfy\sigma}
+\sum_{k\in V^*}\sum_{\lambda=1,2} \omega(k) a(k, \lambda)^*a(k, \lambda)\no
&+ U\sum_{x\in \Lambda} n_{x\uparrow} n_{x\downarrow}+
\sum_{{x, y\in
 \Lambda}\atop{ x\neq y}}U_{xy}n_{x}n_y.
\label{DefHamiltonian}
\end{align}
$\mathsf{H}$ acts in the Hilbert space $\mathfrak{E}_{\mathrm{el}} \otimes
\mathfrak{R}$.
$\mathfrak{R}$ is the Fock space over $\ell^2(V^*\times \{1, 2\})$ with
$V^*=(\frac{2\pi}{L}\BbbZ)^3$, namely,  $\mathfrak{R}=\oplus_{n\ge 0}
\otimes^n_{\mathrm{s}} \ell^2(V^*\times \{1, 2\})$.
$a(k, \lambda)^*$ and $a(k, \lambda)$
 are the bosonic creation- and annihilation operators, respectively. These
 operators satisfy the following commutation relations: 
 \begin{align}
[a(k, \lambda), a(k', \lambda')^*]=\delta_{\lambda\lambda'}
  \delta_{kk'},\ \ 
[a(k, \lambda), a(k', \lambda')]=0.
\end{align} 
The quantized vector potential is given by 
\begin{align}
A(x)=|V|^{-1/2}\sum_{k\in V^*}
 \sum_{\lambda=1,2}\frac{\chi_{\kappa}(k)}{\sqrt{2\omega(k)}}\vepsilon(k,
 \lambda) \Big(
e^{ik\cdot x}a(k, \lambda)+e^{-ik\cdot x}a(k, \lambda)^*
\Big).
\end{align} 
The form factor $\chi_{\kappa}$ is the indicator function of the ball of
radius $0<\kappa<\infty$.
The dispersion relation $\omega(k)$ is chosen to be $\omega(k)=|k|$ for
$k\in V^* \backslash \{0\}$, $\omega(0)=m_0$ with $0<m_0<\infty$.
$C_{xy}$ is a piecewise smooth curve from $x$ to $y$.
For concreteness, the polarization vectors are chosen as  
\begin{align}
\vepsilon(k, 1)
=\frac{(k_2, -k_1, 0)}{\sqrt{k_1^2+k_2^2}}
,\ \  \vepsilon(k, 2)=\frac{k}{|k|} \wedge \vepsilon(k, 1).
\end{align} 
To avoid ambiguity, we set $
\vepsilon(k, \lambda)=0
$ if $k_1=k_2=0$.
 $A(x)$ is essentially self-adjoint. We denote its
closure by the same symbol.
This model was introduced by Giuliani at el. in \cite{GMP}.
Remark that $\mathsf{H}$ is essentially self-adjoint and bounded from below.
We denote its closure by the same symbol.

\begin{rem}
{\rm 
A more precise definition of $\int_{C_{xy}} dr\cdot A(r)$ is given in 
Section \ref{PFRad}. $\diamondsuit$
}
\end{rem} 

\begin{Thm}\label{Radiation1}
Assume {\bf (A. 1)}, {\bf (A. 2)} and {\bf (A. 3)}.
We define the effective Hamiltonian $\mathsf{H}_{\infty}$ by $
\mathsf{H}_{\infty}=
P\mathsf{H}^{U=0}
P$, where   $\mathsf{H}^{U=0}$ is  the Hamiltonian $\mathsf{H}$ with $U=0$.
For all $z\in \BbbC\backslash \BbbR$, we have
\begin{align}
\lim_{U\to \infty}(\mathsf{H}-z)^{-1}=(\mathsf{H}_{\infty}-z)^{-1} P
\end{align} 
 in the operator norm topology.
\end{Thm} 

As before, we denote  the restriction of $\mathsf{H}_{\infty}$ to
$\mathfrak{E}\otimes \mathfrak{R}$ by the same symbol. 

\begin{Thm}\label{Radiation2}
Assume {\bf (A. 1)}, {\bf (A. 2)} and {\bf (A. 3)}.
Assume that $\Lambda$ satisfies the  connectivity condition.
Then the  ground state of $\mathsf{H}_{\infty}$ has total spin  $S=(|\Lambda|-1)/2$ and is
 unique apart from  the trivial $(2S+1)$-degeneracy. 
\end{Thm}

\section{Proof of Theorem \ref{HHEff}} \label{Pf1}
\setcounter{equation}{0}
Let $H_1=P^{\perp} HP^{\perp}$ and let $ H_{01}=PHP^{\perp}+P^{\perp}HP$.
Using the fact that $
P(\sum_{x\in \Lambda} n_{x\uparrow} n_{x\downarrow})P=0
$, we have  $H_{\infty}=PHP$. Accordingly,  we have $H=H_{\infty}+H_1+H_{01}$. Moreover,
since $P$ commutes with boson operators, we have 
\begin{align}
H_{01}=PTP^{\perp}+P^{\perp}TP,
\end{align} 
where $T=\sum_{x, y\in \Lambda} \sum_{\sigma\in \{\uparrow,
\downarrow\}}
t_{xy} c_{x\sigma}^*c_{y\sigma}
$. In particular, $H_{01}$ is  a bounded operator.

We denote the restriction of $A$ to $\ran(P^{\perp})$ by  $A\restriction
P^{\perp}$.
\begin{lemm}\label{U1}
Let $E(H_1)=\min \mathrm{spec}(H_1\restriction P^{\perp})$, where $\mathrm{spec}(A)$ is
 spectrum of $A$. Then 
\begin{align}
E(H_1) \ge C+U,
\end{align} 
where $C$ is a  constant  independent of $U$.
\end{lemm} 
{\it Proof.}
For any self-adjoint operator $A$, bounded from below, we set 
$\mathcal{E}(A):=\inf \mathrm{spec}(A)$.
We divide $H_1$ into two parts as 
$
H_1=H_{1, U=0}+U\sum_{x\in \Lambda} n_{x\uparrow} n_{x\downarrow} P^{\perp}
$, where $H_{1, U=0}$ is the Hamiltonian $H_1$ with $U=0$.
We have  
 $
E(H_{1}) \ge \mathcal{E}(H_{1, U=0}) +\mathcal{E}(U
\sum_{x\in \Lambda} n_{x\uparrow} n_{x\downarrow} \restriction P^{\perp}
).$
Since $
\mathcal{E}(U
\sum_{x\in \Lambda} n_{x\uparrow} n_{x\downarrow}\restriction  P^{\perp}
) \ge U
$, we obtain the desired result. $\Box$

\begin{coro}\label{U2}
If $U$ is sufficiently large, then $H^{-1}_1P^{\perp}$ is a bounded operator.
\end{coro}

\begin{lemm}\label{U3}
Let  $z\in \BbbC\backslash \BbbR$.  We have 
\begin{align}
\|(H_1-z)^{-1} P^{\perp}\| \le  \{E(H_1)-|z|\}^{-1},
\end{align} 
provided that $U$ is sufficiently large.
\end{lemm} 
{\it Proof.} Since  $H_1^{-1}P^{\perp}$ is bounded, we have
\begin{align}
(H_1-z)^{-1}P^{\perp}=\sum_{n=0}^{\infty} (H_1^{-1}P^{\perp} z)^n H_1^{-1}P^{\perp}.
\end{align} 
Thus, we have 
$\displaystyle 
\|(H_1-z)^{-1 }P^{\perp}\| \le \sum_{n=0}^{\infty} E(H_1)^{-n-1} |z|^n=\{E(H_1)-|z|\}^{-1}
$. $\Box$

\begin{lemm}\label{U4}
Let $z\in \BbbC\backslash \BbbR$. If $|\mathrm{Im} z|$ is sufficiently
 large, then we have
\begin{align}
 \lim_{U\to \infty}
\Big\|
(H-z)^{-1}-(H_{\infty}+H_1-z)^{-1}
\Big\|=0.
\end{align} 
\end{lemm} 
{\it Proof.} First, remark that 
\begin{align}
\Big\{
(H-z)^{-1}-(H_{\infty}+H_1-z)^{-1}
\Big\}P
=(H-z)^{-1}P^{\perp} (-H_{01}) (H_{\infty}-z)^{-1}P. \label{Exp1}
\end{align} 
The norm of $(H-z)^{-1} P^{\perp}$ is estimated as follows:
Since 
\begin{align}
(H-z)^{-1} P^{\perp}=\sum_{n=0}^{\infty}(-1)^n
 \Big\{(H_{\infty}+H_1-z)^{-1} H_{01}\Big\}^n
(H_1-z)^{-1}P^{\perp},
\end{align} 
we have, by Lemma \ref{U3},
\begin{align}
\|(H-z)^{-1} P^{\perp}\|& \le 
\Bigg(
\sum_{n=0}^{\infty} |\mathrm{Im} z|^{-n} \|H_{01}\|^n
\Bigg) \{E(H_1)-|z|\}^{-1}\no
&:=C_z\{E(H_1)-|z|\}^{-1}.
\end{align} 
Thus, by Lemma \ref{U1} and (\ref{Exp1}),
\begin{align}
\Bigg\| 
\Big\{
(H-z)^{-1}-(H_{\infty}+H_1-z)^{-1}
\Big\}P
\Bigg\| \le C_z\|H_{01}\| |\mathrm{Im} z|^{-1}\{E(H_1)-|z|\}^{-1} \to 0
\end{align} 
as $U\to \infty$.

Next, since 
\begin{align}
\Big\{
(H-z)^{-1}-(H_{\infty}+H_1-z)^{-1}
\Big\}P^{\perp}=(H-z)^{-1} P(-H_{01}) (H_1-z)^{-1} P^{\perp},
\end{align} 
we obtain, by Lemmas \ref{U1} and  \ref{U3},
\begin{align}
\Bigg\|
\Big\{
(H-z)^{-1}-(H_{\infty}+H_1-z)^{-1}
\Big\}P^{\perp}
\Bigg\| \le |\mathrm{Im} z|^{-1}\|H_{01}\| \{E(H_1)-|z|\}^{-1} \to 0
\end{align}  
as $U\to \infty$. $\Box$

\begin{lemm}\label{U5}
Let $z\in \BbbC\backslash \BbbR$. If $|\mathrm{Im} z|$ is sufficiently
 large, then we have
\begin{align}
\lim_{U\to \infty}\Big\|
(H_{\infty}+H_1-z)^{-1}-(H_{\infty}-z)^{-1}P
\Big\|=0.
\end{align} 
\end{lemm} 
{\it Proof.}
Remark that 
\begin{align}
(H_{\infty}+H_1-z)^{-1}-(H_{\infty}-z)^{-1}P=
(H_1-z)^{-1}P^{\perp}.
\end{align} 
Hence, we obtain, by Lemmas \ref{U1} and  \ref{U3},
\begin{align}
\Big\|
(H_{\infty}+H_1-z)^{-1}-(H_{\infty}-z)^{-1}P
\Big\| \le \{E(H_1)-|z|\}^{-1}\to 0
\end{align} 
as $U\to \infty$.  $\Box$

\begin{flushleft}{\it
Completion of  proof of Theorem \ref{HHEff}
}
\end{flushleft} 

By Lemmas \ref{U4}, \ref{U5} and \cite[Theorem VIII. 19]{ReSi1}, we obtain the desired result in the
theorem. $\Box$

\section{Operator theoretic correlation inequalities} \label{Pf2}
\setcounter{equation}{0}
\subsection{Self-dual cones}

In this section, we will review a general theory of correration
inequalities
developed in \cite{Miyao1, Miyao8, Miyao2, Miyao4, Miyao5,Miyao6,
Miyao7, Miyao10}.

Let $\h$ be a complex Hilbert space.
By a {\it convex  cone}, we understand a closed convex set  $\cone\subset \h$
such that $t\cone \subseteq \cone$ for all $t\ge 0$ and $\cone\cap (-\cone)=\{0\}$. In what follows, we always assume that $\cone\neq
\{0\}$.

\begin{define}{\rm
The {\it dual cone   of} $\cone$ is defined by 
\begin{align}
\cone^{\dagger}=\{\eta\in \h\, |\, \la \eta|\xi\ra\ge 0\ \forall \xi\in
\cone \}.
\end{align} 
 We say that $\cone$ is {\it self-dual} if 
$
\cone=\cone^{\dagger}. \ \ \ \ \ \ \diamondsuit
$
}\end{define}

\begin{rem}{\rm \cite{Miyao10}}\label{HilCone}
{\rm
$\cone$ is a self-dual cone if and only if $\cone$ is a Hilbert cone.\footnote{ 
Let $\h$ be a complex Hilbert space.  A convex cone
 $\cone$ in $\h$ is called  a {\it Hilbert cone}, if it
 satisfies
 the following: 
(i) $ \la \xi| \eta\ra\ge 0$ for all $\xi, \eta\in \cone$;
(ii) Let $\h_{\BbbR}$
 be a real subspace of $\h$ generated by $\cone$ . Then
	     for all $\xi\in \h_{\BbbR}$, there exist $\xi_+,
	     \xi_-\in \cone$ such that $\xi=\xi_+-\xi_-$ and $\la \xi_+|
	     \xi_-\ra=0$;
(iii) $\h=\h_{\BbbR}+i
	    \h_{\mathbb{R}}= \{\xi+ i \eta\, |\, \xi, \eta\in
	     \h_{\BbbR}\}$.
} $\diamondsuit$
}
\end{rem}

\begin{define}
{\rm 
\begin{itemize}
\item[(i)] A vector $\xi$ is said to be  {\it positive w.r.t. $\cone$} if $\xi\in
 \cone$.  We write this as $\xi\ge 0$  w.r.t. $\cone$.

 \item[(ii)] A vector $\eta\in \cone$ is called {\it strictly positive
w.r.t. $\cone$} whenever $\la \xi| \eta\ra>0$ for all $\xi\in
\cone\backslash \{0\}$. We write this as $\eta>0 $
w.r.t. $\cone$. $\diamondsuit$

\end{itemize} 
}
\end{define}

\begin{example}\label{Ex1}
{\rm 
Let $\mathfrak{X}$ be a finite-dimensional  Hilbert space and let $\{x_n\}_{n=1}^N$
 be a complete orthonormal system in $\mathfrak{X}$. We set 
\begin{align}
\cone= \mathrm{Coni}\{x_n\, |\, n=1, \dots, N\},
\end{align} 
where $\mathrm{Coni}(S)$ is the conical hull of $S$ . Then $\cone$ is a self-dual cone in $\mathfrak{X}$.
$x\ge 0$ w.r.t. $\cone$  if and only if
 $\la x_n|x\ra\ge 0$ for all $n\in \{1, \dots, N\}$.
In addition, $x>0$ w.r.t. $\cone$ if and only if $\la x_n|x\ra>0$ for
 all $n\in \{1, \dots, N\}$.
 $\diamondsuit$

}
\end{example}
\begin{example}\label{Ex2}
{\rm 
Let $(M, \mu)$ be a $\sigma$-finite measure space. We set 
\begin{align}
L^2(M,  d\mu)_+=\{f\in L^2(M, d\mu)\, |\, f(m)\ge 0\ \ \mbox{$\mu$-a.e.}
\}.
\end{align}
$L^2(M,  d\mu)_+$ is a self-dual cone in $L^2(M, d\mu )$. Clearly, 
$f\ge 0$ w.r.t. $L^2(M, d\mu)_+$
 if and only if $f(m) \ge 0$ $\mu$-a.e.. On the other hand, $f >0$
 w.r.t. $L^2(M, d\mu)_+$ if and only if $f(m)>0$ $\mu$-a.e..
 $\diamondsuit$ 
}
\end{example}

\subsection{Operator inequalities associated with self-dual cones}

In  subsequent  sections, we  use the following operator inequalities.
\begin{define}{\rm 
We denote by  $\mathscr{B}(\h)$  the set of all bounded linear operators on
$\h$.
Let $A, B\in \mathscr{B}(\h)$. Let $\cone$ be a self-dual cone in $\h$.

 If $A \cone\subseteq \cone,$\footnote{
For each subset $\mathfrak{C}\subseteq \h$, $A\mathfrak{C}$ is
	     defined by $A\mathfrak{C}=\{Ax\, |\, x\in \mathfrak{C}\}$.
} we then 
write  this as  $A \unrhd 0$ w.r.t. $\cone$.\footnote{This
 symbol was introduced by Miura \cite{Miura}.} In
	     this case, we say that {\it $A$ preserves the
positivity w.r.t. $\cone$.} 
Let $\mathfrak{H}_{\BbbR}$ be a real subspace generated by $\cone$.
 Suppose that $A\h_{\BbbR}\subseteq
 \h_{\BbbR}$ and $B\h_{\BbbR} \subseteq
	     \h_{\BbbR}$. If $(A-B) \cone\subseteq
	     \cone$, then we write this as $A \unrhd B$ w.r.t. $\cone$. $\diamondsuit$
} 
\end{define} 

\begin{rem}\label{Pequiv}
{\rm
$A\unrhd 0$ w.r.t. $\cone$ $\Longleftrightarrow$ $ \la \xi |A\eta\ra\ge
 0$
for all $\xi, \eta\in \cone$. $\diamondsuit$
}
\end{rem} 

\begin{example}\label{Ex3}
{\rm 
Let $\mathfrak{X}$ be a finite-dimensional Hilbert space and let
 $\mathfrak{P}$ be a self-dual cone given in Example \ref{Ex1}.
A matrix $A$ acting in $\mathfrak{X}$ satisfies $A\unrhd 0$
 w.r.t. $\mathfrak{P}$
 if and only if $\la x_m|Ax_n\ra \ge 0$ for all $m, n\in \{1, \dots,
 N\}$. $\diamondsuit$
}
\end{example} 
{\it Proof.} Let $x, y\in \mathfrak{P}$. We can express $x$ and $y$ as 
$x=\sum_{n=1}^N a_n x_n$ and $y=\sum_{n=1}^Nb_n x_n$ with $a_n, b_n\ge
 0$.
Since $\la x|Ay\ra=\sum_{m, n=1}^Na_mb_n A_{mn}$
,  we conlcude the assertion by Remark \ref{Pequiv}.  $\Box$

\begin{example}\label{Ex4}
{\rm 
Let $L^2(\BbbR)$ be the space of Lebesgue square integrable functions on
 $\BbbR$. Let $p$ be a differential operator defined by $p=-i d/dx$.
For each $a\in \BbbR$, we set $U_a=e^{iap}$. 
The unitary operator $U_a$ satisfies $U_a \unrhd 0$
 w.r.t. $L^2(\BbbR)_+$. $\diamondsuit$
}
\end{example} 
{\it Proof.} Remark that $(U_a f)(x)=f(x+a)$ a.e. $x$ for each $f\in
L^2(\BbbR)$. If $f\in L^2(\BbbR)_+$, that is, $f(x) \ge 0$ a.e. $x$, then 
$(U_a f)(x)=f(x+a) \ge 0$ a.e. $x$. $\Box$ 
\medskip\\

The following proposition is fundamental  in the present paper.

\begin{Prop}\label{BC}
Let $A, B, C, D\in \mathscr{B}(\h)$ and let $a, b\in
 \BbbR$. 
\begin{itemize}
\item[{\rm (i)}] If $A\unrhd 0, B\unrhd 0$ w.r.t. $\cone$ and
	     $a, b\ge 0$, then $aA +bB \unrhd 0$
	     w.r.t. $\cone$.
\item[{\rm (ii)}] If $A \unrhd B \unrhd 0$ and $C\unrhd D \unrhd 0$
	     w.r.t. $\cone$,
	     then
$AC\unrhd BD \unrhd 0$ w.r.t. $\cone$.
\item[{\rm (iii)}] If $A \unrhd 0 $ w.r.t. $\cone$, then $A^*\unrhd 0$ w.r.t. $\cone$.
\end{itemize} 
\end{Prop} 
{\it Proof.} (i) is trivial.

(ii) If $X\unrhd 0$ and $Y\unrhd 0$
w.r.t. $\cone$, 
we have $XY\cone\subseteq X\cone \subseteq \cone$.
Hence,  it holds that $XY\unrhd 0$
w.r.t. $\cone$.
Hence, we have 
\begin{align*}
AC-BD=\underbrace{A}_{\unrhd 0}\underbrace{(C-D)}_{\unrhd 0}+
 \underbrace{(A-B)}_{\unrhd 0} \underbrace{D}_{\unrhd 0} \unrhd 0\ \ \
 \mbox{w.r.t. $\cone$}.
\end{align*} 

(iii) For each $\xi, \eta\in \cone$, we know that 
\begin{align}
\la \xi|A^*\eta\ra=\la \underbrace{A}_{\unrhd 0} \underbrace{\xi}_{\ge 0}|
\underbrace{\eta}_{\ge 0}\ra \ge 0.
\end{align}
Thus, by Remark \ref{Pequiv}, we conclude (iii).  $\Box$
\medskip\\

\begin{Prop}\label{Basic}
Let $\{A_n\}_{n=1}^{\infty}\subseteq \mathscr{B}(\h)$  and let $A\in
 \mathscr{B}(\h)$.  Suppose that $A_n$ converges to $A$ in the weak
 operator topology. If $A_n\unrhd 0$ w.r.t. $\cone$ for all $n\in
 \BbbN$, then $A\unrhd 0$ w.r.t. $\cone$.
\end{Prop} 
{\it Proof.} By Remark \ref{Pequiv}, $\la \xi|A_n\eta\ra\ge 0$
for all $\xi, \eta\in \cone$. Thus, 
$\displaystyle 
\la \xi|A\eta\ra=\lim_{n\to \infty}\la \xi|A_n\eta\ra\ge 0
$ for all $\xi, \eta\in \cone$. By Remark \ref{Pequiv} again, we
conclude that $A\unrhd 0$ w.r.t. $\cone$. $\Box$

\begin{define}
{\rm 
Let $A\in \mathscr{B}(\h)$.
We write  $A\rhd 0$ w.r.t. $\cone$, if  $A\xi >0$ w.r.t. $\cone$ for all $\xi\in
\cone \backslash \{0\}$. 
 In this case, we say that {\it $A$ improves the
positivity w.r.t. $\cone$.} $\diamondsuit$
}
\end{define}

\begin{example} \label{Ex5}
{\rm 
Let $\mathfrak{X}$ and  $\mathfrak{P}$ be given in Example \ref{Ex1}.
Let $A$ be an $N\times N$ matrix acting in $\mathfrak{X}$. 
Suppose that $A \unrhd 0$ w.r.t. $\mathfrak{P}$.
If, for each $m, n\in \BbbN$, there
 exists an $\ell\in \{0\} \cup \BbbN$ such that $\la x_m|A^{\ell}
 x_n\ra>0$, then $e^{A} \rhd 0$ w.r.t. $\mathfrak{P}$. $\diamondsuit$
}
\end{example} 
{\it Proof.} Let $u, v\in \mathfrak{P} \backslash \{0\}$. Then we can
express 
$u$ and $v$ as $
u=\sum_{n=1}^N u_n x_n
$ and $v=\sum_{n=1}^N v_n x_n$
 with $u_n, v_n \ge 0$. Since $u$ and $v$ are non-zero vectors, there
 exist
$m, n\in \{1, \dots, N\}$ such that $u_m>0$ and $v_n>0$. 
Remark that, since $A\unrhd 0$ w.r.t. $\mathfrak{P}$, it holds that
$A^k\unrhd 0$ w.r.t. $\mathfrak{P}$ for all  $k\in \BbbN$.   In particular, all matrix elements of $A^k$ are positive numbers.
By choosing  $\ell\in \{0\} \cup \BbbN$ such that $\la x_m|A^{\ell} x_n\ra>
0$, we have
\begin{align}
\la u |e^A v\ra=\sum_{k=0}^{\infty} \frac{1}{k!} \underbrace{\la
 u|A^kv\ra}_{\ge 0}
\ge \frac{1}{\ell!} \la u |A^{\ell} v\ra \ge \frac{u_m v_n}{ \ell!} \la
 x_m|A^{\ell} x_n\ra>0.
\end{align} 
Thus, $e^A \rhd 0$ w.r.t. $\mathfrak{P}$. $\Box$

\begin{example}\label{Ex6}
{\rm 
Let  $L^2(\BbbR)$ be the space of Lebesgue square integrable functions on
 $\BbbR$. Let us consider an  operator $h=-d^2/dx^2+x^2$  acting  in
 $L^2(\BbbR)$. $h$ is self-adjoint and bounded from below.
Moreover, $e^{-\beta h} \rhd 0$ w.r.t. $L^2(\BbbR)_+$ for all $\beta>0$,
 see, e.g., \cite[Theorems XIII. 44 and XIII. 47]{ReSi4}. $\diamondsuit$
}
\end{example} 

The following theorem  plays  an important role.
\begin{Thm}\label{PFF}{\rm (Perron--Frobenius--Faris)}
Let $A$ be a  self-adjoint  operator, bounded from below. 
We set $E(A):=\inf \mathrm{spec}(A)$.
Suppose that 
 $0\unlhd e^{-tA}$ w.r.t. $\cone$ for all $t\ge 0$  and that  $E(A)$ is an eigenvalue.
Let $P_A$ be the orthogonal projection onto the closed subspace spanned
 by  eigenvectors associated with   $E(A)$.
 Then,  the following
 are equivalent:
\begin{itemize}
\item[{\rm (i)}] 
$\dim \ran P_A=1$ and $P_A\rhd 0$ w.r.t. $\cone$.
\item[{\rm (ii)}] $0\lhd e^{-tA}$ w.r.t. $\cone$ for all
	     $t>0$.
\item[{\rm (iii)}] For each $\xi, \eta\in \cone\backslash\{0\}$,
there exists a $t>0$ such that $\la \xi| e^{-tA} \eta\ra>0$.
\end{itemize}
\end{Thm} 
{\it Proof.} See, e.g.,    \cite{Faris, Miyao1, ReSi4}. $\Box$

\begin{rem}
{\rm 
 (i) is equivalent to the following: there exists a unique $\xi\in \h$ such that $\xi>0$
 w.r.t. $\cone$
and $P_A=|\xi\ra \la \xi|$. 
Of course, $\xi$ satisfies $A\xi=E(A)\xi$.
$\diamondsuit$
}
\end{rem}

\section{Proof of Theorem \ref{Holstein}}\label{Pf3}
\setcounter{equation}{0}

\subsection{The $S^{(3)}=M$ subspace}
Recall the definition of $
S^{(3)}
$ given in (\ref{DSpin}). We denote the spectrum of $S^{(3)}$ by $\mathrm{spec}(S^{(3)})$.
Remark that  $\mathrm{spec}(S^{(3)})
=\{-(|\Lambda|-1)/2, -(|\Lambda|-3)/2, \dots, (|\Lambda|-1)/2\}
$. Thus, we have the following decomposition:
\begin{align}
\mathfrak{E}=\bigoplus_{M\in \mathrm{spec}(S^{(3)})} \mathfrak{E}(M),\ \
 \ 
\mathfrak{E}(M)=\ker(S^{(3)}-M)\cap
 \mathfrak{E}.
\end{align} 
In this paper, we call $\mathfrak{E}(M) \otimes \Fock$ the $S^{(3)}=M$
{\it subspace}.

\subsection{The Lang-Firsov transformation}

Let 
\begin{align}
L=\omega^{-1} \sum_{x, y\in \Lambda} g_{xy}n_x(b_y^*-b_y).
\end{align} 
$L$ is essentially anti-self-adjoint. We denote its closure by $L$.
We set $H'_{\infty}=e^LH_{\infty}e^{-L}$.

Let 
\begin{align}
\mathbf{T}=\sum_{x, y\in \Lambda} \sum_{\sigma=\uparrow, \downarrow} \vartheta_{xy}T_{xy}(\sigma), 
\end{align} 
where 
\begin{align}
T_{xy}(\sigma)=t_{xy} Pc_{x\sigma}^* c_{y\sigma}P
\end{align} 
and  
\begin{align}
\vartheta_{xy}=\exp\Bigg\{
-i \sqrt{2} \omega^{-3/2} \sum_{z\in \Lambda}
(g_{xz}-g_{yz})p_{z}
\Bigg\}
\end{align} 
with $p_x$ the self-adjoint extension of $i\sqrt{\frac{\omega}{2}}(b_x^*-b_x)$.

Since $P$ commutes with $L$, we obtain 
\begin{align}
H_{\infty}'=\mathbf{T}+\Ue P+\omega N_{\mathrm{b}} P+\mathrm{Const.}, \label{EffHami}
\end{align} 
where $N_{\mathrm{b}}=\sum_{x\in \Lambda} b_x^*b_x$ and 
\begin{align}
\Ue=\sum_{x\neq y} U_{\mathrm{eff}, xy} n_xn_y
\end{align} 
with $
U_{\mathrm{eff}, xy}=U_{xy}-\omega^{-1}\sum_{z\in \Lambda} g_{xz}g_{zy}.
$
Henceforth, we ignore the constant term in (\ref{EffHami}), because it
does not affect our proof below.

\subsection{Definition of the connector and its related operators }

For each $(x, \bsig) \in \mathcal{C}_M$, we define
$\bsig'=\{\sigma_z'\}_{z\in \Lambda}\in \{\uparrow,
\downarrow\}^{\Lambda}$ by 
\begin{align}
\sigma_z'=\begin{cases}
\uparrow &\mbox{ if $z=x$}\\
\sigma_z & \mbox{otherwise}.
 \end{cases} 
\end{align} 
Following Tasaki \cite{Tasaki1, Tasaki2}, we define a complete orthonormal system $\{|x,
\bsig\ra\, |\, (x, \bsig) \in \mathcal{C}_M\}\subset \mathfrak{E}$ by 
\begin{align}
|x, \bsig\ra=c_{x\uparrow} \prod'_{z\in \Lambda} c_{z\sigma_z'}^*\Omega_{\mathrm{f}},
\end{align} 
where
$\Omega_{\mathrm{f}}$ is the Fock vacuum and 
 $\prod\rq{}_{z\in \Lambda}$ indicates the ordered product according to an arbitrarily fixed order in $\Lambda$.

For each $M\in \{-(|\Lambda|-1)/2, \ -(|\Lambda|-3)/2, \dots, (|\Lambda|-1)/2\}$, 
a canonical self-dual cone in $\mathfrak{E}(M)$ is defined by 
\begin{align}
\mathfrak{E}_+(M)=\mathrm{Coni}\Big\{
|x,\bsig\ra\, \big|\, (x, \bsig)\in \mathcal{C}_M
\Big\}.
\end{align}

We begin with the following lemma.

\begin{lemm}\label{txy}
For every $x,  y\in \Lambda$ and $\sigma\in \{\uparrow, \downarrow\}$,
we have
\begin{align}
\la y',\bsig|\{-T_{xy}(\sigma)\} |x', \btau\ra
=t_{xy}\delta_{xx'} \delta_{yy'} \delta_{S_{yx}(\btau)\bsig}. 
\label{Tconn}
\end{align} In additon, 
 $-T_{xy}(\sigma) \unrhd 0$
 w.r.t. $\mathfrak{E}_+(M)$. 

\end{lemm} 
{\it Proof.}
To check (\ref{Tconn}) is easy, or  see \cite[Eq. (4.3)]{Tasaki2}.
Let $\phi, \psi\in \mathfrak{E}_+(M)$. Thus, we can express these as 
$
\phi=\sum_{(x, \bsig) \in \mathcal{C}_M} \phi_{x, \bsig}|x, \bsig\ra,\ \ 
\psi=\sum_{(x, \bsig) \in \mathcal{C}_M} \psi_{x, \bsig}|x, \bsig\ra
$ with $\phi_{x, \bsig} \ge 0$ and $\psi_{x, \bsig} \ge 0$. 
Using these expressions and (\ref{Tconn}), we can check that
$\la \phi|\{-T_{xy}(\sigma)\} \psi\ra\ge0$. 
Thus, by 
Example \ref{Ex3}, we conclude that 
$-T_{xy}(\sigma) \unrhd 0$
 w.r.t. $\mathfrak{E}_+(M)$.  $\Box$

\begin{define}{\rm 
We say that 
$p=
\{x_1, x_2, \dots, x_{\ell}
\}$ in Definition \ref{HC} is a {\it connector } between
$(x, \bsig)$ and  $(y, \btau)$.
The number $\ell-1$ is called the {\it length of $p$} and denoted by
 $|p|$. $\diamondsuit$

For each connector $p$ between $(x, \bsig)$ and $(y, \btau)$, we define a
linear operator $\tau(p)$ by 
\begin{align}
\tau(p) =  \{-T_{x_{\ell-1}
 x_{\ell}}(\sigma_{\ell})\}
 \{-T_{x_{\ell-2}
 x_{\ell-1}}(\sigma_{\ell-1})\}
\cdots
\{- T_{x_{1}
 x_{2}}(\sigma_{1})\}.
\end{align} 
}
\end{define}

\begin{Prop}\label{ConnT}
Let $(x, \bsig), (y, \btau)\in \mathcal{C}_M$. Let $p$ be a connector
 between $(x, \bsig)$ and $(y, \btau)$. Then we have 
\begin{align}
\tau(p) \unrhd t_{x_{\ell}x_{\ell-1}} \cdots t_{x_2x_1} |y, \btau 
\ra
\la
x,\bsig|\ \ \mbox{ w.r.t. $\mathfrak{E}_+(M)$}.
\end{align}
In particular, we have
$
\la y, \btau|\tau(p)|x, \bsig\ra\ge t_{x_{\ell} x_{\ell-1}} \cdots t_{x_2x_1}>0.
$
\end{Prop} 
{\it Proof.} We set 
\begin{align}
(x_1, \bsig_1):=(x, \bsig),\ (x_2,
\bsig_2):=S_{x_2x_1}(x_1, \bsig_1),\dots, (x_{\ell},
\bsig_{\ell})
:=S_{x_{\ell} x_{\ell-1}}(x_{\ell-1}, \bsig_{\ell-1}),
\end{align} 
where $S_{xy}$ is defined by (\ref{DefSxy}).
Remark that $(x_{\ell}, \bsig_{\ell})=(y, \btau)$.
Define 
\begin{align}
P_1:=|x_1, \bsig_1\ra\la x_1, \bsig_1|,\dots, P_{\ell}:=|x_{\ell},
\bsig_{\ell}\ra\la x_{\ell}, \bsig_{\ell}|.
\end{align} 
Since $\one \unrhd P_k$ w.r.t. $\mathfrak{E}_+(M)$ for all $k=1, \dots,
\ell$,\footnote{
To show this, remark that $
\one =\sum_{(x, \bsig) \in \mathcal{C}_M} |x, \bsig\ra\la x, \bsig|
$. Since $|x, \bsig\ra\la x, \bsig| \unrhd 0$ w.r.t. $\mathfrak{E}_+(M)$
 for all $(x, \bsig) \in \mathcal{C}_M$, we obtain that  $\one \unrhd |x,
 \bsig\ra\la x, \bsig| $ w.r.t. $\mathfrak{E}_+(M)$ for all $(x, \bsig)
 \in \mathcal{C}_M$.
} 
we obtain, by Proposition \ref{BC} and Lemma \ref{txy},
\begin{align}
\tau(p)& = \one
 \big\{-T_{x_{\ell-1}
 x_{\ell}}(\sigma_{\ell})
\big\}\one
 \big\{
-T_{x_{\ell-2}x_{\ell-1}}(\sigma_{\ell-1})
\big\} \one
 \cdots
\one
\big\{
- T_{x_{1}
 x_{2}}(\sigma_{1})
\big\}\one\no
 &\unrhd 
P_{\ell}\big\{
- T_{x_{\ell-1}
 x_{\ell}}(\sigma_{\ell})
\big\}
P_{\ell-1}
 \big\{
-T_{x_{\ell-2}x_{\ell-1}}(\sigma_{\ell-1})
\big\}
 P_{\ell-2}
 \cdots
P_{2}
\big\{
- T_{x_{1}
 x_{2}}(\sigma_{1})
\big\}
P_1\no
& \unrhd 
t_{x_{\ell}x_{\ell-1}} \cdots t_{x_2x_1} |y, \btau 
\ra
\la
x,\bsig|.
\end{align} 
This completes the proof. $\Box$

\begin{lemm}\label{UP}
For each $\beta \ge 0$, $e^{-\beta \Ue P} \unrhd 0$
 w.r.t. $\mathfrak{E}_+(M)$
 for all $\beta \ge 0$.
\end{lemm} 
{\it Proof.} For each $(x, \bsig)\in \mathcal{C}_M$, $|x, \bsig\ra $ 
is an eigenvector of $\Ue$. 
We denote the corresponding eigenvalue by $\Ue(x, \bsig)$.
Then one sees that 
$
e^{-\beta \Ue} |x,\bsig\ra=e^{-\beta \Ue(x, \bsig)}|x,\bsig\ra
$. Since $e^{-\beta \Ue(x, \bsig)}$ is a positive number, we conclude
the assertion by Remark \ref{Pequiv}. $\Box$

\subsection{The Schr\"odinger representation}
Let  $q_x=
\frac{1}{\sqrt{2\omega}}(b_x+b_x^*)
$.  $q_x$ is  essentially self-adjoint.
We denote its closure by the same symbol.
Recall that $p_x$ is the  closure of $i \sqrt{\frac{\omega}{2}}(b_x^*-b_x)$. 

The bosonic Fock space $\Fock$ can be naturally identified with $L^2(Q)$
with $Q=\BbbR^{|\Lambda|}$. In addition, 
$q_x$ and $p_x$ can be
identified with a multiplication operator and $p_x=-i \partial/\partial
q_x$, respectively. In what follows, we use this representation.
A natrual self-dual cone in $L^2(Q)$ is 
\begin{align}
L^2(Q)_+=
\big\{f\in L^2(Q)\, \big|\, f(\bq) \ge 0\  \mbox{a.e. } \big\}.
\end{align}

\begin{lemm}\label{PhP}
We have the following:
\begin{itemize}
\item[{\rm (i)}] For each $x, y\in \Lambda$, $\vartheta_{xy} \unrhd
	     0$ w.r.t. $L^2(Q)_+$.
\item[{\rm (ii)}] $e^{-\beta \omega N_{\mathrm{b}}} \rhd 0$
	     w.r.t. $L^2(Q)_+$
for all $\beta >0$.
\end{itemize} 
\end{lemm} 
{\it Proof.} 
(i) Since $p_x=-i \partial/\partial q_x$, $e^{i a  p_x}$ is a translation
operator. Thus, by Example \ref{Ex4},  $e^{ia p_x} \unrhd 0$ w.r.t. $L^2(Q)_+$ for all $a\in \BbbR$,
which implies $\theta_{xy} \unrhd 0$ w.r.t. $L^2(Q)_+$ for all $x, y\in
\Lambda$.

(ii) We remark that $
\omega N_{\mathrm{b}}=\frac{1}{2}\sum_{x\in \Lambda}
(-\partial^2/\partial q_x^2+q_x^2)
-\frac{|\Lambda|}{2}$. Hence, (ii) follows from \cite[Theorems XIII. 44
and XIII. 47]{ReSi4} (see also Example \ref{Ex6}).
$\Box$

 \subsection{A natural self-dual cone in $\mathfrak{E}(M)\otimes \Fock$}
First,  we remark the following identification: 
\begin{align}
\mathfrak{E}(M) \otimes L^2(Q)=\int_Q^{\oplus} \mathfrak{E}(M)d\bq.
\end{align} 
We take  the following self-dual cone in the full space $
\mathfrak{E}(M)\otimes \Fock
$:
\begin{align}
\Cone&=\Big\{
\Phi \in \mathfrak{E}(M)\otimes L^2(Q)\, \Big|\, \Phi(\bq) \ge 0 \
 \mbox{w.r.t. $\mathfrak{E}_+(M)$ for a.e. $\bq$}
\Big\}.
\end{align}

\begin{lemm}
$\Cone$ is a self-dual cone in $\mathfrak{E}(M)\otimes L^2(Q)$.
\end{lemm} 
{\it Proof.}
We will prove that $\Cone^{\dagger}=\Cone$. It is easy to check that 
$
\Cone \subseteq \Cone^{\dagger}
$ . Thus, it suffices to show the converse.
Let $\Psi=\int_{Q}^{\oplus} \Psi(\bq) d\bq \in \Cone^{\dagger}$. Then, for all
$\Phi\in \Cone$, we have $\la \Psi|\Phi\ra\ge 0$.
By choosing $\Phi$ as $\Phi=|x, \bsig\ra\otimes f$ with $f\in L^2(Q)_+$, 
we have $
\int_Q \la \Psi(\bq)|x, \bsig\ra f(\bq) d\bq \ge 0
$. Since $f$ is arbitrarily, we have $\la \Psi(\bq)|x, \bsig\ra\ge 0$
for all $(x, \bsig)\in \mathcal{C}_M$, which implies that $
\psi(\bq) \ge 0
$ w.r.t. $\mathfrak{E}_+(M)$. Hence, $\Psi\in \Cone$. $\Box$

\begin{lemm}\label{TensorP}
Let $A\in \mathscr{B}(\mathfrak{E}(M))$ and $B\in \mathscr{B}(L^2(Q))$.
Assume that $A\unrhd 0$ w.r.t. $\mathfrak{E}_+(M)$ and $B\unrhd 0$
 w.r.t. $L^2(Q)_+$. We have 
$
A\otimes \one \unrhd 0,\ \one \otimes B\unrhd 0
$ and $A\otimes B \unrhd 0$ w.r.t. $\Cone$.

\end{lemm} 
{\it Proof.} Let $\Psi=\int^{\oplus}_Q \Psi(\bq) d\bq \in \Cone$.
Thus, $\Psi(\bq) \ge 0$ w.r.t. $\mathfrak{E}_+(M)$ for a.e. $\bq$.
Remark that 
\begin{align}
A\otimes \one \Psi=\int_Q^{\oplus} A\Psi(\bq) d\bq. \label{DirectI}
\end{align} 
Since $A\unrhd 0$ w.r.t. $\mathfrak{E}_+(M)$, it holds that $A\Psi(\bq)
\ge 0$ w.r.t. $\mathfrak{E}_+(M)$ for a.e. $\bq$. Thus, the RHS of
(\ref{DirectI}) is positive w.r.t. $\Cone$. This implies that  $A\otimes \one
\unrhd 0$ w.r.t. $\Cone$.

We decompose $\Psi$ as $\Psi=\sum_{(x, \bsig) \in \mathcal{C}_M}
|x, \bsig\ra\otimes f_{x, \bsig}
$, where $f_{x, \bsig}(\bq)=\la x, \bsig|\Psi(\bq)\ra$.
Since $\Psi\ge 0$ w.r.t. $\Cone$, $f_{x, \bsig}(\bq) \ge 0$ for
a.e. $\bq$ and  every $(x, \bsig) \in \mathcal{C}_M$.
Because $B\unrhd 0$ w.r.t. $L^2(Q)_+$, we have that 
$Bf_{x, \bsig} \ge 0$ w.r.t. $L^2(Q)_+$, which implies
$
\one \otimes B\Psi=\sum_{(x, \bsig) \in \mathcal{C}_M} |x,
\bsig\ra\otimes Bf_{x, \bsig} \ge 0
$ w.r.t. $\Cone$. Thus, we conclude that $\one \otimes B \unrhd 0$
w.r.t. $\Cone$.

Finally, we have $A\otimes B=(A\otimes \one )(\one \otimes B) \unrhd 0$
w.r.t. $\Cone$. $\Box$

\begin{rem}
{\rm 
Let $A\in \mathscr{B}(\mathfrak{E}(M))$ and $B\in \mathscr{B}(L^2(Q))$.
In this paper,  we abbrevicate the tensor products $A\otimes \one $ and
 $\one \otimes B$ simply as $A$ and $B$ if no confusion arises.
For example, a claim that $A\unrhd 0$ w.r.t. $\Cone$ means that $
A\otimes \one \unrhd 0$ w.r.t. $\Cone$. $\diamondsuit$
}
\end{rem}

\begin{Prop}\label{ConnectP}
Let $p$ be a connector between $(x, \bsig)$ and $(y, \btau)$.
We have 
$
(-\mathbf{T})^{|p|} \unrhd  \vartheta_{yx}\tau(p)\unrhd 0
$
w.r.t. $\Cone$.
\end{Prop} 
{\it Proof.} By Lemmas \ref{txy}, \ref{PhP} and \ref{TensorP}, we have 
\begin{align}
-\mathbf{T}\unrhd \theta_{xy}\{-T_{xy} (\sigma)\} \unrhd 0 \label{Hoshi}
\end{align} 
 w.r.t. $\Cone$ for all $x,y\in \Lambda$ and $\sigma\in \{\uparrow,
\downarrow\}$. By using this and Proposition \ref{BC} (ii), we have
\begin{align}
(-\mathbf{T})^{\ell-1} \unrhd \big\{
-\theta_{x_{\ell-1} x_{\ell}} T_{x_{\ell-1} x_{\ell}}(\sigma_{\ell})
\big\}
\cdots \big\{
-\theta_{x_{1} x_{2}} T_{x_{1} x_{2}}(\sigma_{\ell})
\big\} \label{TInq}
\end{align} 
w.r.t. $\Cone$. Since $
\theta_{x_1x_2}\cdots
\theta_{x_{\ell-1} x_{\ell}}  =\theta_{x_1x_{\ell}}
=\theta_{xy}
$, we obtain the desired result. $\Box$

\subsection{Uniqueness of the ground states}

The purpose of this subsection is to prove the following:

\begin{Thm}\label{PISemi}
For each $
M\in \{-(|\Lambda|-1)/2,\ -(|\Lambda|-3)/2, \dots, (|\Lambda|-1)/2\}
$,  it holds that $e^{-\beta H\rq{}_{\infty}} \rhd 0$ w.r.t. $\Cone$ for all $\beta >0$.
Thus, the ground states of $H_{\infty}\rq{}$ is unique
 and can be chosen to be  strictly positivie w.r.t. $\Cone$
 by Theorem \ref{PFF}.

\end{Thm} 

To prove Theorem \ref{PISemi},
we begin with the following lemma.
\begin{lemm}
$e^{-\beta H_{\infty}'} \unrhd 0$ w.r.t. $\Cone$ for all $\beta \ge 0$.
\end{lemm} 
{\it Proof.}
By (\ref{Hoshi}) and Proposition \ref{BC} (ii),  we have  $(-\mathbf{T})^n \unrhd 0$ w.r.t. $\Cone$ for
all $n\in \BbbN$, which implies that 
\begin{align}
e^{-\beta \mathbf{T}} =\sum_{n=0}^{\infty} 
\underbrace{\frac{\beta^n}{n!}}_{\ge 0}
\underbrace{
(-\mathbf{T})^n
}_{\unrhd 0} \unrhd 0
\end{align} 
w.r.t. $\Cone$ for all $\beta \ge 0$. On the other hand, we have,
by Lemmas \ref{UP}, \ref{PhP} and \ref{TensorP},
\begin{align}
e^{-\beta (\Ue P+\omega N_{\mathrm{b}})} =
\underbrace{
e^{-\beta \Ue P}
}_{\unrhd  0}
\underbrace{
 e^{-\beta \omega N_{\mathrm{b}} }
}_{\unrhd 0} \unrhd 0
\end{align} 
w.r.t. $\Cone$ for all $\beta \ge 0$.
Thus, by the Trotter-Kato product formula and Proposition \ref{Basic}, we have 
\begin{align}
e^{-\beta H_{\infty}'} =\mbox{strong-}\lim_{n\to \infty} 
\Big(
e^{-\beta \mathbf{T}/n} e^{-\beta (\Ue P+\omega N_{\mathrm{b}})/n}
\Big)^n \unrhd 0
\end{align} 
w.r.t. $\Cone$ for all $\beta\ge 0$, because 
$
e^{-\beta \mathbf{T}/n}e^{-\beta(\Ue P+\omega N_{\mathrm{b}})} \unrhd 0
$ for all $\beta \ge 0$ and $n\in \BbbN$. $\Box$
\medskip\\

Let $H_{\infty, 0}'$ be the Hamiltonian $H_{\infty}'$ with $U_{xy}=0$
for all $x, y\in \Lambda$.

\begin{Prop}\label{Equiv}
The following {\rm (i)} and {\rm (ii)} are equivalent:
\begin{itemize}
\item[{\rm (i)}] $e^{-\beta H_{\infty}'} \rhd 0$ w.r.t. $\Cone$ for all
	     $\beta >0$.
\item[{\rm (ii)}] $e^{-\beta H_{\infty, 0}'} \rhd 0$ w.r.t. $\Cone$ for all
	     $\beta >0$.

\end{itemize} 
\end{Prop} 
{\it Proof.} To prove Proposition \ref{Equiv}, it suffices to check the
conditions (a)-(c) in Theorem \ref{PEq}.
To this end, we set $
U_{\mathrm{eff}}^{(n)}=(1-\frac{1}{n}) \Ue
$. Let $H_{\infty}'^{(n)}=H_{\infty, 0}'+U_{\mathrm{eff}}^{(n)}P$.
Choose  $\psi \in \mathfrak{E}[M] \hat{\otimes} \Fock_0$, arbitrarily, where 
$\Fock_0$ is the finite particle subspace of $\Fock$ \footnote{
To be precise, 
\begin{align}
\Fock_0=\{\Phi=\{\Phi_n\}_{n=0}^{\infty}\, |\, \mbox{There exists an
 $n_0\in \{0\} \cup \BbbN$ such that $\Phi_n=0$ for all $n\ge n_0$}\}.
\end{align} 
} and $\hat{\otimes }$ indicates the algebraic tensor product.
We easily check that 
$
H_{\infty}'^{(n)} \psi\to H_{\infty}'\psi,\ \
(H_{\infty}'-U_{\mathrm{eff}}^{(n)}P) \psi \to H_{\infty, 0}'\psi
$ as $n\to \infty$. Thus, by \cite[Theorem VIII. 25]{ReSi1}, $H_{\infty}'^{(n)}$
 converges to $H_{\infty}'$ and $H_{\infty}'-U_{\mathrm{eff}}^{(n)} P$
 converges
to $H_{\infty, 0}'$ in the strong resolvent sense\footnote{
Let $A$ be a self-adjoint operator and let $\{A_n\}_{n=1}^{\infty}$  be a family of self-adjoint
operators.
Then $A_n$ is said to converge to $A$ in the {\it strong resolvent sense} if 
$(A_n-\lambda)^{-1}  \to (A-\lambda)^{-1} $ 
for all $\lambda$ with $\mathrm{Im} \lambda\neq 0$ in the strong
operator topology.

}.
Thus, the condition (a) is satisfied.
To check (b) is easy.

To prove (c), let $\vphi, \psi\in \Cone$ with $\la
\vphi|\psi\ra=0$. Hence, 
\begin{align}
\sum_{(x, \bsig) \in \mathcal{C}_M} \la \vphi_{x, \bsig}|\psi_{x, \bsig}\ra_{L^2(Q)}=0.
\end{align} 
Since $\vphi_{x, \bsig}(\bq) \ge 0$ and $\psi_{x, \bsig}(\bq) \ge 0$  for
a.e. $\bq$,  it holds that $
\la \vphi_{x,\bsig}|\psi_{x, \bsig}\ra =0
$ for all $(x, \bsig) \in \mathcal{ C}_M$.
Recalling the notations in the proof of Lemma \ref{UP}, we have 
\begin{align}
\la \vphi|e^{-\beta U_{\mathrm{eff}}^{(n)}P} \psi\ra=\sum_{(x, \bsig) \in
 \mathcal{C}_M}e^{-\beta (1-n^{-1}) \Ue(x, \bsig)} \la \vphi_{x,\bsig}|\psi_{x, \bsig}\ra =0.
\end{align} 
Thus, (c) is satisfied.  $\Box$
\medskip\\

For our purpose, it suffices to prove that $e^{-\beta H_{\infty, 0}'}
\rhd 0$ w.r.t. $\beta > 0$ by Proposition \ref{Equiv}.
To this end, we use the Duhamel expansion:
\begin{align}
e^{-\beta H_{\infty, 0}'}=\sum_{n=0}^{\infty} D_n(\beta), \label{Duha}
\end{align} 
where 
\begin{align}
D_n(\beta )=\int_{0\le s_1\le \cdots \le s_n \le \beta}
\{-\mathbf{T}(s_1)\}\cdots \{-\mathbf{T}(s_n)\} e^{-\beta \omega
 N_{\mathrm{b}}} \label{DefD}
\end{align} 
with $
\mathbf{T}(s)=e^{-s\omega  N_{\mathrm{b}}} \mathbf{T} e^{s\omega N_{\mathrm{b}}}
$.  Note that the RHS
 of (\ref{Duha}) converges in the operator norm topology.

\begin{lemm}\label{PosiD}
For all $n\in \BbbN_0:=\{0\} \cup \BbbN$ and $\beta \ge 0$, $D_n(\beta)\unrhd 0$ w.r.t. $\Cone$.
\end{lemm} 
{\it Proof.} By Lemma \ref{PhP} and (\ref{Hoshi}),  the integrand in the RHS of
(\ref{DefD})
satisfies
\begin{align}
\underbrace{e^{-s_1 \omega N_{\mathrm{b}}}
}_{\unrhd 0}
\underbrace{
 (-\mathbf{T})
}_{\unrhd 0}
\underbrace{
 e^{-(s_2-s_1)\omega
 N_{\mathrm{b}} } 
}_{\unrhd 0}
\cdots 
\underbrace{
e^{-(s_n-s_{n-1}) \omega N_{\mathrm{b}}}
}_{\unrhd 0}
\underbrace{
 (-\mathbf{T})
}_{\unrhd 0}
\underbrace{
e^{-(\beta-s_n) \omega N_{\mathrm{b}}}
}_{\unrhd 0}
\unrhd 0
\end{align} 
w.r.t. $\Cone$ for all $0 \le s_1\le \cdots \le s_n\le \beta$.
Thus, it holds that  $D_{n}(\beta) \unrhd 0$
w.r.t. $\Cone$. $\Box$

\begin{define}\label{Ergodic}
{\rm 
We say that 
$\{D_n(\beta)\}$ is {\it ergodic} w.r.t. $\Cone$ if, for each $\vphi, \psi\in
 \Cone\backslash \{0\}$, there exist $n\in \BbbN_0$ and $\beta \ge 0$ such
 that $\la \vphi|D_n(\beta ) \psi\ra>0$. $\diamondsuit$
}
\end{define}

\begin{lemm}
If $\{D_n(\beta)\}$ is ergodic w.r.t. $\Cone$, then $e^{-\beta
 H_{\infty, 0}'} \rhd 0$ w.r.t. $\Cone$ for all $\beta >0$.
\end{lemm} 
{\it Proof.}
Choose $\vphi, \psi\in \Cone\backslash \{0\}$, arbitrarily.
By the ergodicity of $\{D_n(\beta)\}$, there exist $n_0\in \BbbN_0$ and
$\beta_0\ge 0$ such that $\la \vphi|D_{n_0}(\beta_0) \psi\ra >0$.
By Lemma \ref{PosiD}, we have 
$
e^{-\beta_0 H_{\infty, 0}'} \unrhd D_{n_0}(\beta_0)
$ w.r.t. $\Cone$, which implies 
$
\la \vphi|e^{-\beta_0 H_{\infty, 0}'}\psi\ra\ge \la \vphi|D_{n_0}(\beta_0) \psi\ra>0 
$. By Theorem \ref{PFF}, we conclude that $e^{-\beta  H_{\infty, 0}'} \rhd
0$
w.r.t. $\Cone$ for all $\beta >0$. $\Box$
\medskip\\

For each $n\in \BbbN$ and $\beta \ge 0$, we set
\begin{align}
C_n(\beta)=(-\mathbf{T})^n e^{-\beta  \omega N_{\mathrm{b}}}.
\end{align} 
We remark that, 
in a way similar to Definition \ref{Ergodic}, the ergodicity of
$\{C_{n}(\beta)\}$ can be   defined. 
\begin{lemm}
If $\{C_n(\beta)\}$ is ergodic w.r.t. $\Cone$, then $\{D_n(\beta)\}$ is
 ergodic w.r.t. $\Cone$.
\end{lemm} 
{\it Proof.} We denote the integrand of the RHS of (\ref{DefD}) by
$F(s_1, \dots, s_n)$.
In the proof of Lemma \ref{PosiD}, we have already proved that $
F(s_1, \dots, s_n) \unrhd 0$ w.r.t. $\Cone$, provided that 
$0\le s_1\le \cdots \le s_n \le \beta$.
Also remark that $F(0, \dots, 0)=C_n(\beta)$.

Let $\vphi, \psi \in \Cone \backslash \{0\}$.
Since $\{C_n(\beta)\}$ is ergodic, there exist $n_0\in \BbbN_0$ and
$\beta_0\ge 0$ scuh that $\la \vphi|C_{n_0}(\beta_0) \psi\ra>0$.
Thus, since  a function $\la \vphi|F(s_1, \dots, s_n)\psi\ra$ is
continuous 
in $s_1, \dots, s_n$, we have 
\begin{align}
\la \vphi|D_{n_0}(\beta_0) \psi\ra =\int_{0\le s_1\le \cdots\le s_n \le
 \beta}
\la \vphi|F(s_1, \dots, s_n)\psi\ra>0.
\end{align} 
This means that $\{D_n(\beta)\}$ is ergodic. $\Box$

\begin{Thm}\label{CERGO}
$\{C_n(\beta)\}$ is ergodic w.r.t. $\Cone$.
\end{Thm} 
{\it Proof.}
In this proof, we will use the following notation:
Let $\xi,\eta\in \Cone$. If $\xi-\eta\in \Cone$, then we express this as 
$\xi \ge \eta$ w.r.t. $\Cone$.

The following property is useful: Let $A$ be a linear operator acting in
$\mathfrak{E}(M)\otimes L^2(Q)$. Suppose that $A\unrhd 0$
w.r.t. $\Cone$.
If $\xi \ge \eta$ and $\xi'\ge \eta'$ w.r.t. $\Cone$, then  we have 
$
\la \xi|A\xi'\ra\ge \la \eta|A\eta'\ra
$.

For each $\vphi, \psi\in \mathfrak{P} \backslash \{0\}$, there exist
 $(x, \bsig)$ and $(y, \btau)$ such that 
\begin{align}
\vphi \ge |x, \bsig\ra\otimes \vphi_{x, \bsig},\ \ \ 
\psi\ge |y, \btau\ra\otimes \psi_{y, \btau}\ \ \ \mbox{w.r.t. $\Cone$},
\end{align} 
where $\vphi_{x, \bsig}, \psi_{y, \btau} \in L^2(Q)_+\backslash \{0\}$.\footnote{
To see this, we remark that $\vphi$ can be expressed as 
$
\vphi=\sum_{(x, \bsig) \in \mathcal{C}_M} |x, \bsig\ra\otimes \vphi_{x, \bsig}
$, where $\vphi_{x, \bsig}(\bq)=\la x, \bsig|\vphi(\bq)\ra$.
Since $\vphi(\bq) \ge 0$ w.r.t. $\mathfrak{E}_+(M)$, it holds that $
\vphi_{x, \bsig} \ge 0
$ w.r.t. $L^2(Q)_+$.
Thus, $|x, \bsig\ra \otimes \vphi_{x, \bsig} \in \Cone$ for all $(x,
\bsig)\in \mathcal{C}_M$, which implies $
\vphi\ge |x, \bsig\ra \otimes \vphi_{x, \bsig}
$ for all $(x, \bsig) \in \mathcal{C}_M$.
Since $\vphi$ is  nonzero, there exists a $(x_0, \bsig_0) \in
\mathcal{C}_M$
such that $\vphi_{x_0, \bsig_0} \neq 0$. Hence, we obtian $\vphi \ge
|x_0, \bsig_0\ra \otimes \vphi_{x_0, \bsig_0}$.
}
By the connectivity condition, there exists a connector  $p$ between  $(x,
\bsig)$ 
and 
$(y, \btau)$. By Proposition \ref{ConnectP}, we have 
\begin{align}
\la \vphi|C_{|p|}(\beta) \psi\ra
\ge 
\la x, \bsig|\tau(p)|y, \btau\ra  \la \vphi_{x, \bsig}|
\theta_{yx}
e^{-\beta
 \omega N_{\mathrm{b}}} \psi_{y, \btau}\ra. \label{Lower1}
\end{align} 
Since $\theta_{yx} \unrhd 0$ and $e^{-\beta \omega N_{\mathrm{b}}} \rhd
0$  w.r.t. $L^2(\mathcal{Q})_+$ for all $\beta >0$, it holds that 
\begin{align}
 \la \vphi_{x, \bsig}|
\theta_{yx}
e^{-\beta
 \omega N_{\mathrm{b}}} \psi_{y, \btau}\ra>0, 
\end{align} 
which implies that the RHS of (\ref{Lower1}) is strictly positive by
Proposition \ref{ConnT}.
Thus, $\{C_n(\beta)\}$ is ergodic. $\Box$
\medskip\\

As a consequence of Theorem \ref{CERGO}, we obtain Theorem
\ref{PISemi}. $\Box$

\subsection{Completion of proof of Theorem \ref{Holstein}}
Recall the definitions of 
$
S^{(-)}
$ and $ S^{(+)}$ given in (\ref{DefSpin}).
Remark that $
S^{(-)}
$ maps $\mathfrak{E}(M)$ into $\mathfrak{E}(M-1)$ for all $M\in
\mathrm{spec}(S^{(3)}) \backslash \{-(|\Lambda|-1)/2\}$.
Moreover, we have the following:

\begin{lemm}\label{SPP}
$S^{(-)}$ maps $\mathfrak{E}_+(M) \backslash \{0\}$ into 
$\mathfrak{E}_+(M-1) \backslash \{0\}$ for all $M\in
 \mathrm{spec}(S^{(3)})\backslash\{-(|\Lambda|-1)/2\}
$.
\end{lemm} 
{\it Proof.} We just remark that 
\begin{align}
S^{(-)}|x, \bsig\ra=\sum_{y\in \Lambda,\ x\neq y} |x, t_y(\bsig)\ra,
\end{align} 
where 
\begin{align}
\big(t_y(\bsig)\big)_z=\begin{cases}
\sigma_z & z\neq y\\
\overline{\sigma}_z & z=y
\end{cases} 
\end{align} 
with $\overline{\uparrow}=\downarrow$ and
$\overline{\downarrow}=\uparrow$. $\Box$
\medskip\\

By Theorem  \ref{PISemi}, the ground states of $H_{\infty}'$ in the
$S^{(3)}=M$ subspace is unique. We denote the unique ground state by $\psi_M$.
Remark that  $\psi_M>0$ w.r.t. $\Cone$.
We set $\psi^{\dagger}=\psi_{M=\frac{1}{2}(|\Lambda|-1)}$.
$\psi^{\dagger}$ can be expressed as 
\begin{align}
\psi^{\dagger}=\sum_{x\in \Lambda} |x, \Up_x\ra\otimes f_x,
\end{align} 
where $f_x>0$ w.r.t. $L^2(Q)_+$ and the spin configuration $\Up_x$ is
defined by 
\begin{align}
(\Up_x)_y=\begin{cases}
\uparrow & y\neq x\\
0 & y=x
\end{cases}. 
\end{align}  
By Lemma \ref{SPP}, $
(S^{(-)})^n \psi^{\dagger} \ge 0
$ w.r.t. $\mathfrak{P}_{M=\frac{1}{2}(|\Lambda|-n-1)}$
and $(S^{(-)})^n \psi^{\dagger} \neq 0$ whenever $0 \le n \le 2(|\Lambda|-1)$.
Let $E_M$ be the ground state energy of $H_{\infty}'$ in the $S^{(3)}=M$
subspace. We set $E^{\dagger}=E_{M=\frac{1}{2}(|\Lambda|-1)}$.
Since $S^{(-)}$ commutes with $H_{\infty}'$, we have 
$
H_{\infty}'(S^{(-)})^n \psi^{\dagger}=E^{\dagger} (S^{(-)})^n \psi^{\dagger}
$. 
Because $\psi_M$ is strictly positive w.r.t. $\Cone$, we have 
$
\la \psi_M|(S^{(-)})^{|\Lambda|-2M-1} \psi^{\dagger}\ra>0
$. Hence, $(S^{(-)})^{|\Lambda|-2M-1} \psi^{\dagger}$ is the ground
state of $H_{\infty}'$ in the $S^{(3)}=M$ subspace and $E^{\dagger}=E_M$
for all $M\in \mathrm{spec}(S^{(3)})$.
Accordingly, $E^{\dagger}$ is the lowest eigenvalue of $H_{\infty}'$
 and $\psi^{\dagger}$ is the ground state of $H_{\infty}'$.
Clearly, $\psi^{\dagger}$ has total spin $S=\frac{1}{2}(|\Lambda|-1)$.
Since $E_M=E^{\dagger}$ for all $
M\in \mathrm{spec}(S^{(3)})
$, every  $\psi_M$
are ground states of $H_{\infty}'$ as well. This completes the proof of
Theorem \ref{Holstein}.
 $\Box$

\section{Proof of Theorems \ref{Radiation1} and \ref{Radiation2}} \label{PFRad}

\setcounter{equation}{0}
\subsection{A remark on the Hamiltonian}
We  have to clarify  a mathematical definition of the integral $
\int_{C_{xy}} dr\cdot A(r)$.
For each $x, y\in \Lambda$ with $x\neq y$, let
\begin{align}
F_{x, y}(k)=\frac{1}{ik\cdot (y-x)} (
e^{ik\cdot y}-e^{i k\cdot x}
).
\end{align} 
Remark that $|F_{x, y}(k)| \le 1$ for every $k\in V^*\backslash \{0\}$. We define a field
operator
$\phi$ by 
\begin{align}
\phi=\sum_{k\in V^*} \sum_{\lambda=1, 2} \frac{\chi_{\kappa}(k)}{\sqrt{\omega(k)}} 
\vepsilon(k, \lambda)\cdot \frac{x-y}{|x-y|}
\Big\{
F_{x, y}(k)a(k, \lambda)+\mathrm{h.c.}
\Big\}.
\end{align} 
$\phi$ is essentially self-adjoint on $\mathfrak{E}(M)\hat{\otimes }
\mathfrak{R}_0$. 
Here, $\mathfrak{R}_0$ is the finite particle subspace \footnote{
To be precise,
\begin{align}
\mathfrak{R}_0=
\Big\{
\Phi=\{\Phi_n\}_{n=0}^{\infty}\in \mathfrak{R}\, \Big|\, \mbox{There
 exists an $n_0\in \BbbN_0$ such that, if $n\ge n_0$, then $\Phi_n=0$}
\Big\}.
\end{align} 
}.
By choosing  $C_{xy}$ as $C_{xy}=\{(1-s)x+sy\in V\, |\, s\in [0, 1]\}$,
we easily check that $
\int_{C_{xy}} dr\cdot A(r)=\phi
$ on $\mathfrak{E}(M) \hat{\otimes } \mathfrak{R}_0$.
Thus, $\int_{C_{xy}} dr\cdot A(r)$ is essentially self-adjoint
on $\mathfrak{E}(M) \hat{\otimes } \mathfrak{R}_0$.
We denote its closure by the same symbol.

\subsection{Sketch of the proof}
Since the proof of Theorem \ref{Radiation1} is similar to that of
Theorem \ref{HHEff}, we omit it.
We provide a sketch of a proof of Theorem \ref{Radiation2}.

We switch to the $\mathscr{Q}$-representation \cite{JHB, Miyao11}.
In the $\mathscr{Q}$-representation, $\mathfrak{R}$ is identified with $L^2(\mathscr{Q},
d\mu)$, where $d\mu$ is some Gaussian measure.
$A(x)$ can be regarded as a multiplication operator by some real valued function.

For notational convenience, we introduce two operators:
\begin{align}
\Hf=\sum_{k\in V^*}\sum_{\lambda=1,2} \omega(k) a(k, \lambda)^* a(k,
 \lambda),
\ \ \ 
 \Nf=\sum_{k\in V^*}\sum_{\lambda=1,2}  a(k, \lambda)^* a(k,
 \lambda).
\end{align} 
$\Hf $ and $\Nf$ are essentially self-adjoint. So, we denote their
closures by the same symbols, respectively.

Let $\xi=e^{i\pi \Nf/2}$. Let
$
E(x)=\xi A(x)\xi^{-1}
$. Then we have the following:
\begin{lemm}\label{FieldS}
\begin{itemize}\item[{\rm (i)}] $e^{i \mathbf{a}\cdot  E(x)} \unrhd 0$ w.r.t. $L^2(\mathscr{Q},
	     d\mu)_+$ for all $\mathbf{a}\in \BbbR^3$ and $x\in V$.
\item[{\rm (ii)}] $e^{-\beta \Hf} \rhd 0$ w.r.t. $L^2(\mathscr{Q},
	     d\mu)_+$
for all $\beta >0$.
\end{itemize} 
\end{lemm} 
{\it Proof.} Proofs of (i) and (ii) are similar to those of \cite[Lemma
7.27]{JHB} and \cite[Proposition 7.28]{JHB}, respectively. $\Box$
\begin{rem}
{\rm 
To understand the meaning of  (i), we recall the following example:
Let $L^2(\BbbR)$ be the space of Lebesgue square integrable functions on
 $\BbbR$. For each $a\in \BbbR$, let $V_a$ be  a multiplication operator
 defined by  $(V_a f)(x)=e^{iax} f(x)$. 
$V_a$ is a unitary operator acting in $L^2(\BbbR)$.
 Trivially, $V_a$ does not preserve the positivity
 w.r.t. $L^2(\BbbR)_+$. Consider the Fourier transformation:
\begin{align}
(\mathcal{F} f)(k)=(2\pi)^{-1/2} \int_{\BbbR} f(x) e^{-ikx} dx,\ \ f\in L^2(\BbbR).
\end{align} 
$\mathcal{F}$ is a unitary operator on $L^2(\BbbR)$. As is well-known,
 we have $
\mathcal{F} x\mathcal{F}^{-1}=-id/dx
$. Thus, $
\mathcal{F} V_a \mathcal{F}^{-1}
$
becomes a translation, that is, 
$\mathcal{F} V_a \mathcal{F}^{-1} =\exp(ia p)$, where $p=-i d/dx$.
We already know that $\exp(ia p) \unrhd 0$ w.r.t. $L^2(\BbbR)_+$ by
 Example \ref{Ex4}.

Now, let go back to  Lemma \ref{FieldS}.
Notice  the following correspondence:
\begin{align}
A(x) \leftrightarrow x,\ \ \  E(x) \leftrightarrow - i\frac{d}{dx},\ \ \ \ 
 \xi\leftrightarrow \mathcal{F}.
\end{align} 
In this context, $\xi$ can be regarded as a kind of  Fourier
 transformation
on $L^2(\mathscr{Q}, d\mu)$.
A multiplication operator $e^{i \mathbf{a}\cdot A(x)}$ does not preserve  the 
positivity, but $e^{i\mathbf{a}\cdot E(x)}=\xi e^{i\mathbf{a} \cdot A(x)} \xi^{-1}$
does; namely, $e^{i \mathbf{a}\cdot E(x)}$ becomes a kind of translation in
 $L^2(\mathscr{Q}, d\mu)$.
  $\diamondsuit$
}
\end{rem}

For each $x, y\in \Lambda$, we define 
\begin{align}
\Phi_{xy}=\exp\Bigg\{
i  \int_{C_{\bfx\bfy}} d\bfr\cdot  E(\bfr)\Bigg\}.
\end{align} 

\begin{lemm}
For each $x, y\in \Lambda$, we have $
\Phi_{xy} \unrhd 0
$ w.r.t. $L^2(\mathscr{Q}, d\mu)_+$.
\end{lemm} 
{\it Proof.} For each $N\in \BbbN$, we set 
\begin{align}
\mathscr{A}_N^{x, y}=\sum_{j=1}^{N+1} \frac{1}{N} \frac{x-y}{|x-y|}
 \cdot A
\Big(
x+\frac{j-1}{N}(y-x)
\Big).
\end{align} 
$\mathscr{A}_N^{x, y}$ is essentially self-adjoint. We denote its closure by the same symbol.
Using the formulas $\|(e^{iB}-1)\vphi \| \le \|B\vphi\|$ for  $B$ self-adjoint
and $
\|a(f)^{\#} \vphi\| \le \|f\| \|(N_{\mathrm{f}}+\one)\vphi\|
$, we have 
\begin{align}
\Big\|
\Big(e^{i  \int_{C_{xy}}dr\cdot A(r)}-e^{i  \mathscr{A}_N^{{x, y}}}
\Big)\vphi
\Big\|
\le  2 \Bigg\|
\frac{\chi_{\kappa}}{\sqrt{2\omega}} \vepsilon \cdot \frac{x-y}{|x-y|}
(F_{x,y}-F_{x,y}^{N})
\Bigg\|\|(N_{\mathrm{f}}+\one )\vphi\| \label{Norm}
\end{align} 
for all $\vphi\in \mathfrak{E}(M) \hat{\otimes } \mathfrak{R}_0$, where
$F_{x, y}^N(k)=
\sum_{j=1}^{N+1} \frac{1}{N} \exp\{i k\cdot (x+\frac{j-1}{N}(y-x))\}
$.  
Since $F_{x, y}^N$ converges to $F_{x, y}$ in $\ell^2(V^*\times \{1, 2\})$,
$e^{i  \mathscr{A}_N^{x, y}}$ strongly converges to $e^{i 
\int_{C_{xy}} dr \cdot A(r)
}$ on $
\mathfrak{E}(M) \hat{\otimes } \mathfrak{R}_0
$ by (\ref{Norm}). Since $
\mathfrak{E}(M) \hat{\otimes } \mathfrak{R}_0
$ is dense in
$\mathfrak{E}(M) \otimes \mathfrak{R}$,  the convergence holds on whole  $
\mathfrak{E}(M) \otimes \mathfrak{R}
$.

Since 
\begin{align}
\xi e^{i \mathscr{A}_N^{x, y}} \xi^{-1}
=\prod_{j=1}^{N+1} \exp
\Bigg\{
i\frac{1}{N} \frac{x-y}{|x-y|}\cdot  E\Big(
x+\frac{j-1}{N}(y-x)
\Big)
\Bigg\},
\end{align} 
we have, by Lemma \ref{FieldS} (i),  $\xi e^{i 
\mathscr{A}_N^{x,y}}\xi^{-1}
\unrhd 0$ w.r.t. $L^2(\mathscr{Q}, d\mu)_+$.
Because $\xi e^{i  \mathscr{A}_N^{x, y}} \xi^{-1}$
strongly converges to $e^{i \int_{C_{xy}} dr\cdot E(r)}$, we conclude
the desired result by Proposition \ref{Basic}. $\Box$
\medskip\\

Let 
\begin{align}
\mathbb{T}=\sum_{x, y\in \Lambda} \sum_{\sigma=\uparrow, \downarrow} \Phi_{xy}T_{xy}(\sigma).
\end{align} 
We define a self-dual cone $\mathfrak{D}_M$ by 
\begin{align}
\mathfrak{D}_M=
\Big\{
\Psi\in \mathfrak{E}(M) \otimes L^2(\mathscr{Q}, d\mu)\, \Big|\,
 \Psi(\bphi)
\ge 0\ \mbox{w.r.t. $\mathfrak{E}_+(M)$ for $\mu$-a.e.}
\Big\}.
\end{align} 
Here, we use the following identification:
\begin{align}
\mathfrak{E}(M) \otimes L^2(\mathscr{Q},
 d\mu)=\int^{\oplus}_{\mathscr{Q}} \mathfrak{E}(M) d\mu.
\end{align} 

Corresponding to Proposition \ref{ConnectP}, we have the following: 

\begin{Prop}\label{RadConn}
Let $p$ be a connector between $(x, \bsig)$ and $(y, \btau)$.
 We have 
$
(-\mathbb{T})^{|p|} \unrhd  \Phi_{yx}\tau(p)\unrhd 0
$
w.r.t. $\mathfrak{D}_M$.
\end{Prop}

Let 
\begin{align}
\mathbb{C}_n(\beta)=(-\mathbb{T})^n e^{-\beta \Hf}.
\end{align} 
In a similar way as in the proof of Theorem \ref{PISemi}, we can prove the following:

\begin{lemm}
If $\{\mathbb{C}_n(\beta)\}$ is ergodic w.r.t. $\mathfrak{D}_M$, then $
\xi e^{-\beta \mathsf{H}_{\infty}} \xi^{-1}
 \rhd 0$ w.r.t. $\mathfrak{D}_M$ for all $\beta >0$.
\end{lemm} 
Thus, it suffices to prove that
 $\{\mathbb{C}_n(\beta)\}$ is ergodic w.r.t. $\mathfrak{D}_M$.

For each $\vphi, \psi\in \mathfrak{D}_M \backslash \{0\}$, there exist
 $(x, \bsig)$ and $(y, \btau)$ such that 
\begin{align}
\vphi \ge |x, \bsig\ra\otimes \vphi_{x, \bsig},\ \ \ 
\psi\ge |y, \btau\ra\otimes \psi_{y, \btau},
\end{align} 
where $\vphi_{x, \bsig}, \psi_{y, \btau} \in L^2(\mathscr{Q}, d\mu)_+\backslash \{0\}$.
By the connectivity condition, there exists a connector  $p$ between  $(x,
\bsig)$  and $(y, \btau)$. By Proposition \ref{RadConn}, we have 
\begin{align}
\la \vphi|\mathbb{C}_{|p|}(\beta) \psi\ra
\ge 
\la x, \bsig|\tau(p)|y, \btau\ra  \la \vphi_{x, \bsig}|
\Phi_{yx}
e^{-\beta
 \Hf} \psi_{y, \btau}\ra. \label{Lower2}
\end{align} 
Since $\Phi_{yx} \unrhd 0$ and $e^{-\beta \Hf} \rhd 0$
w.r.t. $L^2(\mathscr{Q}, d\mu)_+$ by Lemma \ref{FieldS}, we obtain 
\begin{align}
\la \vphi_{x,\bsig}|\Phi_{yx} e^{-\beta \Hf} \psi_{y, \btau}\ra>0
,
\end{align} 
 which implies that the RHS of (\ref{Lower2}) is strictly positive by
Proposition \ref{ConnT}. $\Box$

\begin{flushleft}
{\bf Acknowledgments.}
This work was partially supported by KAKENHI (20554421) and
 KAKENHI(16H03942).
I would be grateful to the anoymous referee for useful comments.
\end{flushleft}

\appendix

\section{A fundamental theorem} \label{Tools}

\setcounter{equation}{0}

\begin{Thm}\label{PEq}
Let $A$ and $B$ be self-adjoint operators,  bounded from below.
Assume the following conditions:
\begin{itemize}
\item[{\rm (a)}] There exists a sequence of bounded self-adjoint
	     operator $C_n$ such that $
A+C_n
$ converges to $B$ in the strong resolvent sense and $B-C_n$
converges to $A$ in the strong resolvent sense as $n\to \infty$;
\item[{\rm (b)}] $ e^{-tC_n}\unrhd 0$ w.r.t. $\cone$ for all $t\in \BbbR$
	     and $n\in \BbbN$;
\item[{\rm (c)}] 
For all $\xi, \eta\in \cone$ such that $\la \xi|\eta\ra=0$, it holds
	     that $\la \xi|e^{-tC_n}\eta\ra=0$ for all $n\in \BbbN$ and
	     $t\ge 0$.
\end{itemize} 
The following {\rm (i)} and {\rm (ii)} are mutually equivalent:
\begin{itemize}
\item[{\rm (i)}] $e^{-tA} \rhd 0$ w.r.t. $\cone$ for all $t>0$;
\item[{\rm (ii)}] $e^{-tB} \rhd 0$ w.r.t. $\cone$ for all $t>0$.
\end{itemize} 

\end{Thm} 
{\it Proof.} 
The proof is similar to \cite[Theorem 3]{Faris}.
For readers' convenience, we provide a proof. 

(i) $\Longrightarrow$ (ii): By (a) and the Trotter-Kato formula,
we have
\begin{align}
e^{-tB} =\lim_{n\to \infty}\lim_{k\to \infty}
\Big(e^{-tA/k}
e^{-tC_n/k}
\Big)^k
\end{align} 
in the strong opertor  topology. Since $e^{-tA} \unrhd 0$ w.r.t. $\cone$ for all
$t\ge 0$ and (b), we conclude that $e^{-tB} \unrhd 0$ w.r.t. $\cone$ for
all $t\ge 0$ by Proposition \ref{Basic}.

Let $\xi\in \cone\backslash \{0\}$. We set 
$
K(\xi)=\{\eta\in \cone\, |\, \la \eta| e^{-tB} \xi\ra=0\ \forall t\ge 0\}
$. To prove (ii), it suffices to show that $K(\xi)=\{0\}$.
If $\eta\in \cone$ and $\la \eta|e^{-tB}\xi\ra=0$, then, by (c), 
$
\la e^{sC_n}\eta|e^{-tB}\xi\ra=0
$. Hence, $e^{sC_n} K(\xi) \subseteq K(\xi)$ for all $s\in \BbbR$.
It is trivial that $e^{-tB}K(\xi)\subseteq K(\xi)$ for all $t\ge 0$. Accordingly, 
$
(e^{-tB/k} e^{tC_n/k})^k K(\xi) \subseteq K(\xi)
$, which implies $e^{-tA}K(\xi)\subseteq K(\xi)$
by the Trotter-Kato product formula.
In particular, if $\eta\in K(\xi)$, then $\la \eta|e^{-tA}\xi\ra=0$.
Since $e^{-tA} \rhd 0$ w.r.t. $\cone$ for all $t>0$, we conclude that $\eta=0$.

Similarly, we can prove that  (ii) $\Longrightarrow$ (i). $\Box$

\end{document}